\documentclass[aps,prd,a4paper,reprint,nofootinbib,superscriptaddress,floatfix,preprintnumbers]{revtex4-1}

\usepackage{hyperref}
\usepackage{amsmath}
\usepackage{amsfonts}
\usepackage{amssymb}
\usepackage{color}
\usepackage[utf8]{inputenc}
\usepackage{graphicx}
\usepackage{microtype}
\usepackage{siunitx}
\usepackage{soul}

\newcommand{\Fermi}{\textit{Fermi}}

\begin{document}

\begin{flushleft}
LAPTH-030/20
\end{flushleft}

\title{Investigating the detection of dark matter subhalos as extended sources with \Fermi-LAT}
\date{\today}

\author{Mattia Di Mauro }\email{dimauro.mattia@gmail.com}
\affiliation{NASA Goddard Space Flight Center, Greenbelt, MD 20771, USA}
\affiliation{The Catholic University of America, Department of Physics, Washington DC 20064, USA}
\author{Martin Stref}\email{stref@lapth.cnrs.fr}
\affiliation{Univ.~Grenoble Alpes, USMB, CNRS, LAPTh, F-74000 Annecy, France}
\author{Francesca Calore}\email{calore@lapth.cnrs.fr}
\affiliation{Univ.~Grenoble Alpes, USMB, CNRS, LAPTh, F-74000 Annecy, France}

\begin{abstract}
Cold dark matter (DM) models for structure formation predict that DM subhalos are present in the Galaxy.
In the standard paradigm of DM as weakly interacting massive particle, subhalos are expected to
shine in gamma rays and to provide a signal detectable with current instruments, notably with the Large Area Telescope (LAT) aboard the \Fermi~satellite.  This is the main motivation behind searches for DM signals towards dwarf spheroidal galaxies and unidentified \Fermi-LAT sources. 
A significant angular extension detected from unassociated sources located at relatively high latitudes is considered a ``smoking gun" signature for identifying DM subhalos.
In the present work, we systematically explore, by means of state-of-the-art models of cold DM halos in the Galaxy, the detectability of \textsl{extended} subhalos with \Fermi-LAT.
We simulate a DM signal exploring different assumptions of subhalos distribution in the Galaxy and DM profile, and reconstruct its flux through a realistic \Fermi-LAT analysis pipeline. In the most optimistic case, we show that a detection of \textsl{extended} DM subhalos can be made for  annihilation cross sections higher than $3 \times 10^{-26}$ cm$^3$/s (for a 100 GeV DM mass), still compatible with existing gamma-ray constraints, and that, in this case, the preference for extension of the source (vs point-like hypothesis) is significant. 
For fainter signals, instead, halos not only do not show significant extension, but they are not even detectable significantly as point-like sources.
\end{abstract}

\maketitle
\section{Introduction}\label{intro}
Unveiling the nature of dark matter (DM) remains one of the major challenges for 
particle physics and cosmology.
Despite the achievements on the theoretical and experimental side, the standard paradigm of DM
being made of weakly interacting massive particles (WIMP)~\cite{Bertone:2004pz} is  challenged by null results of detection of these elusive particles with current experiments. 
In particular, indirect detection of DM signals with high-energy photons strongly constrains the WIMP DM parameter space~\cite{Bringmann:2012ez}.

Since they are expected to have a low astrophysical background and are predicted to be dynamically dominated by DM, dwarf spheroidal galaxies represent promising targets for DM identification~\cite{Strigari:2018utn}. 
Besides, DM halos that cannot form stars are predicted to exist in cold DM scenarios of structure formation. While such objects are ``dark" for optical telescopes, gamma-ray instruments may 
unveil DM signals emitted therein. 
Searches for DM in subhalos, either in the faintest detectable dwarf galaxies or in their “dark halos” counterparts, represent a powerful test for the WIMP paradigm.

Typically, searches towards known dwarf galaxies, as well as searches for DM subhalos have been performed under the assumption that the emitted DM signal is point-like (with respect to the angular resolution of the instrument). 
Both data-driven, e.g.~\cite{Calore:2018sdx,GeringerSameth:2011iw}, and template-based, e.g.~\cite{Ackermann:2015zua}, searches towards known dwarf spheroidal galaxies look for excess(es) of photons above the atrophysical background, compatible with point-like DM signal(s) from the dwarf(s) direction(s).
Analogously, most of the sensitivity predictions for DM subhalo detectability~\cite{Calore:2016ogv,Hutten:2019tew,Calore:2019lks}, as well as searches for DM subhalos in unidentified sources~\cite{,Coronado-Blazquez:2019pny,Coronado-Blazquez:2020dyl}, treat DM subhalos as point-like objects.

Nevertheless, DM subhalos may have a significant angular extension in the sky, depending on their position and mass profile. The detection of angular extension of unidentified high-energy gamma-ray sources located at latitudes $|b|>20^{\circ}$ has been advocated to be a ``smoking gun" signature of DM subhalos~\cite{Bertoni:2016hoh}, and it has recently received more and more attention in the literature. In particular, analyses of angular extension of unidentified sources detected by the Large Area Telescope (LAT), aboard the \Fermi~satellite, were performed without much success: No extended halo was found (or finally confirmed) around \Fermi-LAT unidentified sources~\cite{2012JCAP...11..050Z,Bertoni:2015mla,Biteau:2018tmv,Coronado-Blazquez:2019pny}.
Also a complementary approach, looking for optical counterparts of \Fermi-LAT extended and unidentified sources in GAIA data, gave null results~\cite{Ciuca:2018vsz}.
Also, recent sensitivity predictions for DM subhalo identification with future gamma-ray instruments included a spatial analysis of DM halos~\cite{Egorov:2018cip,Chou:2017wrw}.

Although searches for extension in real data have been carried out, to the best of our knowledge there is a lack of a complete analysis of the detectability of angular extension of DM subhalos. 
It is not clear, for example, if the \Fermi-LAT has the capability of detecting subhalos as extended and, if yes, for which DM particle physics parameters (notably, the annihilation cross section) -- there is indeed no a priori reason why the \Fermi-LAT sensitivity to DM extended subhalos should be the same as for point-like subhalos, as for example derived in~\cite{Calore:2016ogv}.
In the present work, \textsl{for the first time} we address this issue and quantify what is the impact of modeling DM subhalos as fully extended objects.
To this end, we rely on semi-analytical models for the distribution and statistics of DM subhalos in the Galaxy, which account for Milky-Way dynamical constraints and include tidal effects which the subhalos are subject to when moving in the Galactic gravitational potential~\cite{Stref:2016uzb}.
Such models do not distinguish between ``dark subhalos" and optically detectable ones (i.e.~dwarf galaxies), since they do not implement any recipe for galaxy formation. 
In what follows, we therefore indicate as ``subhalo" every DM substructure present in the Galaxy.
Our final goal is to quantify the sensitivity of \Fermi-LAT to the brightest \textsl{extended} DM subhalo, and, ultimately, understand how to use cold DM predictions to identify DM subhalo candidates in unidentified sources exploiting angular information.

In Sec.~\ref{sec:model}, we describe models and statistics of the Galactic subhalo population. In particular, we stress the importance of a correlation between intensity of the predicted DM signal and angular extension of DM subhalos. 
In Sec.~\ref{sec:datamock}, we illustrate the setup to simulate \Fermi-LAT data and the analysis 
detection pipeline we follow. We present our results in Sec.~\ref{sec:results}, and conclude in Sec.~\ref{sec:conclusions}.

\section{Subhalo models and statistics}
\label{sec:model}
In this section, we describe the DM subhalo population models we use, as well as the mock subhalo catalogs generated from these models.

\subsection{The subhalo models}
Our analysis is based on the semi-analytical subhalo model developed by Stref \& Lavalle \cite{Stref:2016uzb} which is referred to as SL17 from now on. 
SL17 is built upon the realistic Milky-Way mass model developed by McMillan \cite{2017MNRAS.465...76M} in which the Galactic dark halo is assumed to have a Navarro-Frenk-White (NFW) \cite{Navarro:1995iw} density profile shape. 
Cold DM subhalos are expected to have cuspy density profiles, and the profile shape can be chosen freely in the model. In the following we consider either NFW subhalos or Einasto subhalos (with $\alpha_{\rm Ein}=0.16$ based on \cite{Wang:2019ftp}).
Subhalos are subject to tidal effects as they orbit in the gravitational potential of the Galaxy and its DM halo. Two distinct effects are accounted for in SL17: The tidal mass loss due to the smooth gravitational potential of the Galaxy, and the effect of gravitational shocking experienced by a subhalo crossing the Galactic stellar disk. 
Both these effects strongly impact the subhalo population by stripping off mass from these objects, sometimes destroying them completely. 
The efficiency of this destruction is still a matter of debate. Studies based on cosmological simulations find that subhalos are efficiently disrupted in the inner parts of the Galaxy \cite{Diemand:2008in,Springel2008}. On the other hand, recent semi-analytical studies find that cuspy subhalos such as those predicted by cold DM are very resilient to tides and can survive considerable mass losses \cite{vandenBosch:2017ynq,2020MNRAS.491.4591E}. Also, the disruption observed in cosmological simulations could be due to numerical artifacts \cite{vandenBosch:2018tyt}.
Whether subhalos can be disrupted or not has consequences on predictions for DM searches, in particular indirect searches for self-annihilating DM because the annihilation rate is very high in cuspy subhalos \cite{galaxies7020065}. In the present work, we remain agnostic about the resilience of subhalos to tides and treat it as a theoretical uncertainty for our predictions. We bracket this uncertainty by considering two extreme configurations of the SL17 model. The ``SL17-fragile" configuration corresponds to what is commonly observed in cosmological simulations, i.e.~subhalos are efficiently disrupted by tides. In the ``SL17-resilient" configuration, on the other hand, subhalos can lose most of their mass but the central cusp almost always survives.  
More precisely, in the SL17-fragile configuration it is assumed that a subhalo is disrupted as soon as its tidal radius is smaller than its scale radius $r_{\rm t}\leqslant r_{\rm s}$. In the SL17-resilient configuration, on the other hand, disruption only takes place if $r_{\rm t}\leqslant 0.01\,r_{\rm s}$.
These configurations were originally defined in \cite{galaxies7020065} which we refer the reader to for additional details.

\subsection{The subhalo mock population}
\label{sec:sim}
The SL17 model gives a statistical description of the Galactic subhalo population. More precisely, it provides a recipe to compute the probability distribution function (PDF) of various subhalo parameters (mass $m_{200}$, concentration $c_{200}$ and position).
This model is fully implemented in the \texttt{CLUMPY} public code \cite{2012CoPhC.183..656C,Bonnivard:2015pia,Hutten:2018aix}, which can be used to generate mock subhalo population catalogs starting from these parameters' PDFs.
Each of these catalogs is therefore a realization of the Galactic subhalo population based on the SL17 model. 
\texttt{CLUMPY} also computes the $J$-factor of each subhalo, i.e.~the integral along the line of sight (l.o.s.) of the DM density squared:
\begin{equation}
    J(\Delta\Omega) = \int_{0}^{\Delta\Omega}\int_{\rm l.o.s.}\rho_{\rm DM}^2\,\mathrm{d}l\,\mathrm{d}\Omega
\label{eq:J}
\end{equation}
where $\rho_{\rm DM}$ is the subhalo mass density, and $\Delta\Omega=2\pi(1-\cos(\theta))$ is the solid angle for a viewing angle $\theta$. The $J$-factor appears in the expression of the gamma-ray flux produced by DM annihilation.
The total $J$-factor, i.e.~the $J$-factor integrated up to the full angular extension of a subhalo (i.e.~its tidal radius), is labeled as $J_{\rm tot}$.

\begin{figure}
    \centering
    \includegraphics[width=0.45\textwidth]{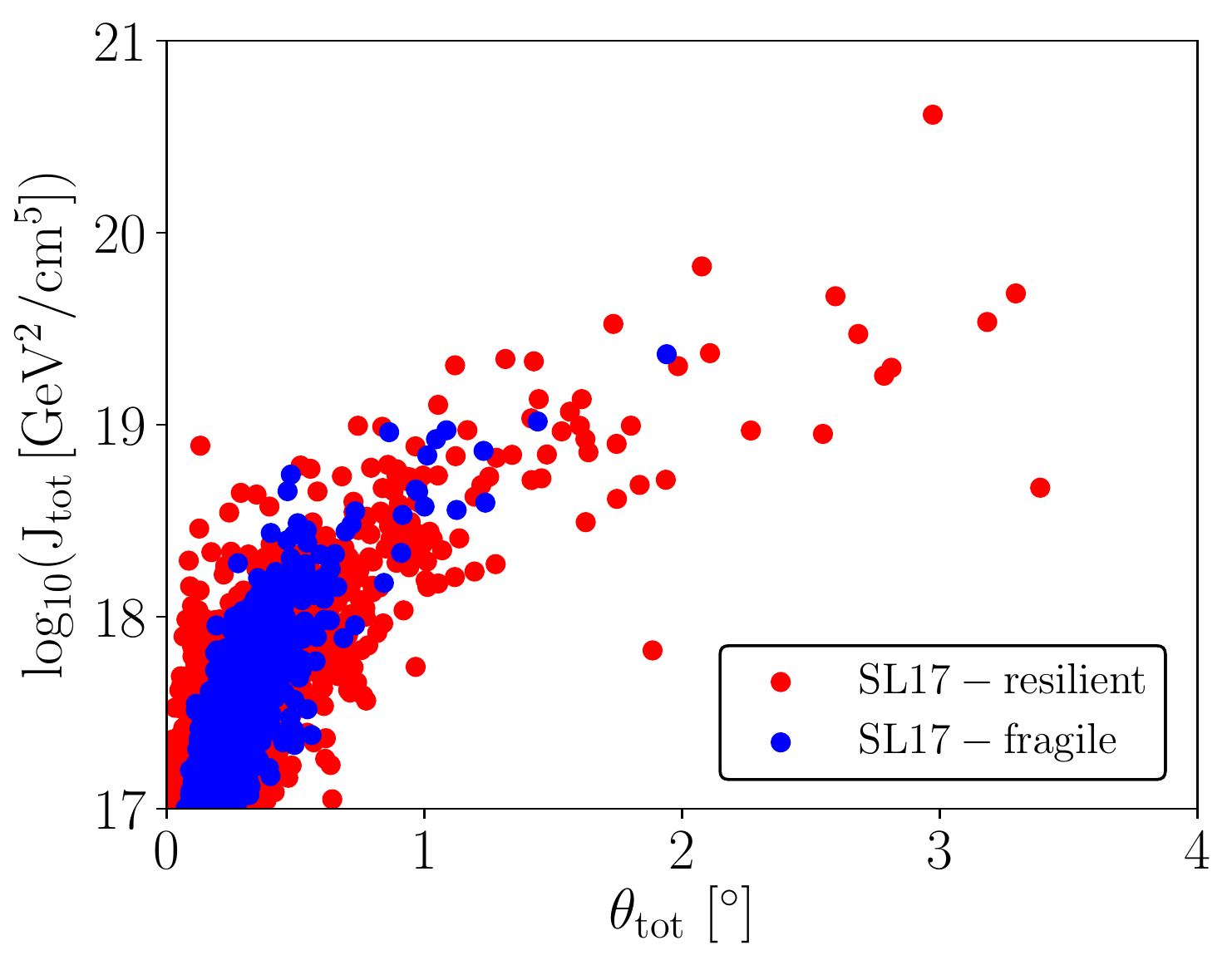}
    \includegraphics[width=0.45\textwidth]{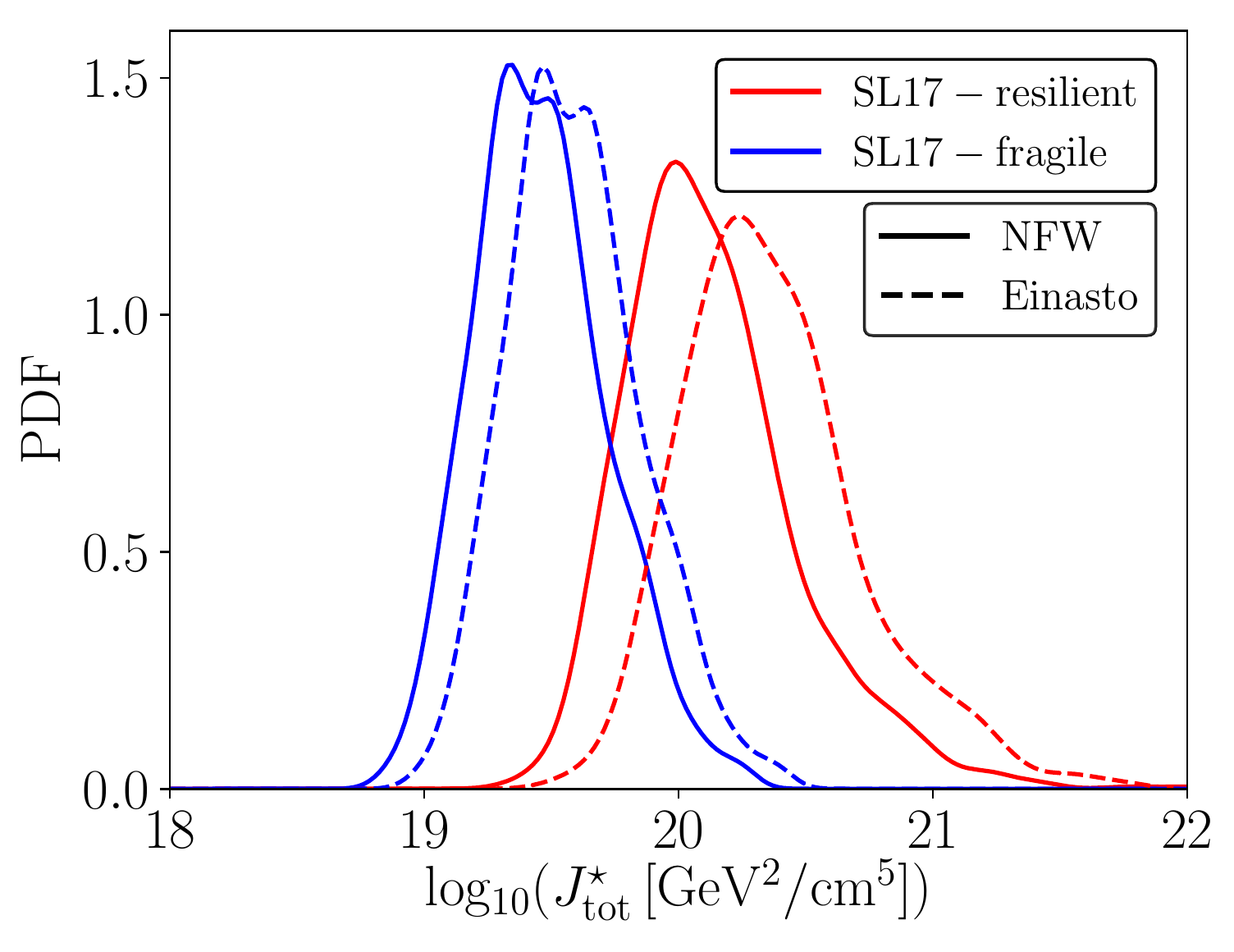}
    \caption{\textbf{Upper panel:} Correlation between $J$-factor and total angular size on the sky for subhalos in two different models (only one realization for each model is shown). \textbf{Lower panel:} $J$-factor PDF of the brightest subhalo, $J_{\rm tot}^\star$.
    }
    \label{fig:corr_jfact_pdf}
\end{figure}

For a subhalo of radius $R$ and at a distance $d$ from the observer, we define the total angular size as: $\theta_{\rm tot} \equiv \arcsin(R/d)$. The angular size of a DM subhalo is therefore a geometric consequence of the subhalo mass profile and its distance.
In the top panel of Fig.~\ref{fig:corr_jfact_pdf}, we show that, in the subhalo catalogs, there exists a correlation between the subhalo $J_{\rm tot}$ and its total angular size on the sky, see also~\cite{Coronado-Blazquez:2019pny}. 
While we are not interested in parameterizing such a correlation nor we directly use it in the following, we can generally conclude that subhalos with the highest $J$-factors also show a significant angular extension of up to a few degrees. This suggests that the DM subhalos with the highest gamma-ray flux could be detected as \textsl{extended} sources rather than point-like objects.
Previous works have mostly focused on the analysis of DM subhalo detectability in the case of point-like sources.
However, if the brightest subhalo is indeed extended in the sky -- as the correlation suggests -- the \Fermi-LAT sensitivity to subhalos may be different.
Here, we aim at quantifying whether or not a search for extended sources improves detection prospects. To do so, we consider the distribution of the brightest subhalo, i.e.~the subhalo with the highest $J$-factor, $J_{\rm tot}^\star$.

We generate 1010 mock population catalogs for each of the two model configurations, SL17-fragile and SL17-resilient. 
We perform a latitude cut in the catalogs, discarding all subhalos with $|b|<10^\circ$ 
(rejecting on average 17\% of subhalos with $J>10^{17}\,\rm GeV^2/cm^5$), and identify, in each Monte Carlo realization, \textsl{the} subhalo with the highest $J$-factor among the remaining ones -- and so only one subhalo for each Monte Carlo realization. 
We then compute the PDF of the $J$-factor of the brightest halo for both configurations and show the result in the bottom panel of Fig.~\ref{fig:corr_jfact_pdf}. If subhalos have a NFW profile, the SL17-fragile PDF peaks around $J_{\rm tot}^\star\sim2.5\times10^{19}\,\rm GeV^2/cm^5$ while the SL17-resilient PDF peaks at $J_{\rm tot}^\star\sim10^{20}\,\rm GeV^2/cm^5$.
The lower $J_{\rm tot}^\star$ in the SL17-fragile case compared to the SL17-resilient case comes mainly from the distance to the brightest object. The stellar disk is very efficient at stripping mass from subhalos. While this is fatal to most clumps passing through the disk in the SL17-fragile scenario, in the SL17-resilient case subhalos can still survive and remain close to the Solar system.
If subhalos have an Einasto profile, the $J$-factors increase by a factor of roughly $1.6$.\footnote{Note that we did not generate subhalo catalogs for the Einasto profile case. Instead we only generate catalogs for the NFW case, find the brightest subhalo in each catalog and extract its parameters (mass $m_{200}$, concentration $c_{200}$ and position), then compute the $J$-factor that would have a subhalo with an Einasto profile with identical parameters.}

\begin{figure}
    \centering
    \includegraphics[width=0.45\textwidth]{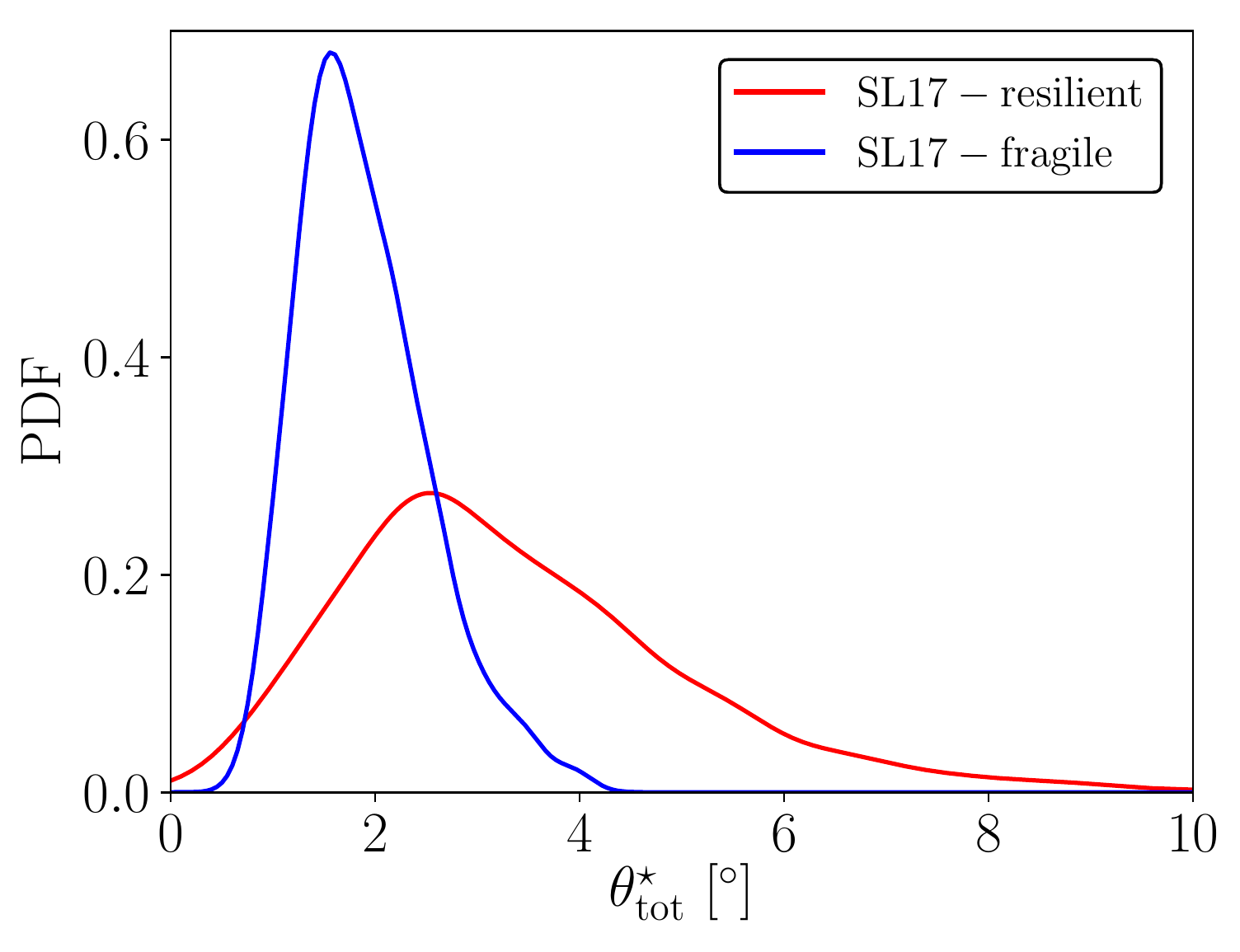}
    \includegraphics[width=0.45\textwidth]{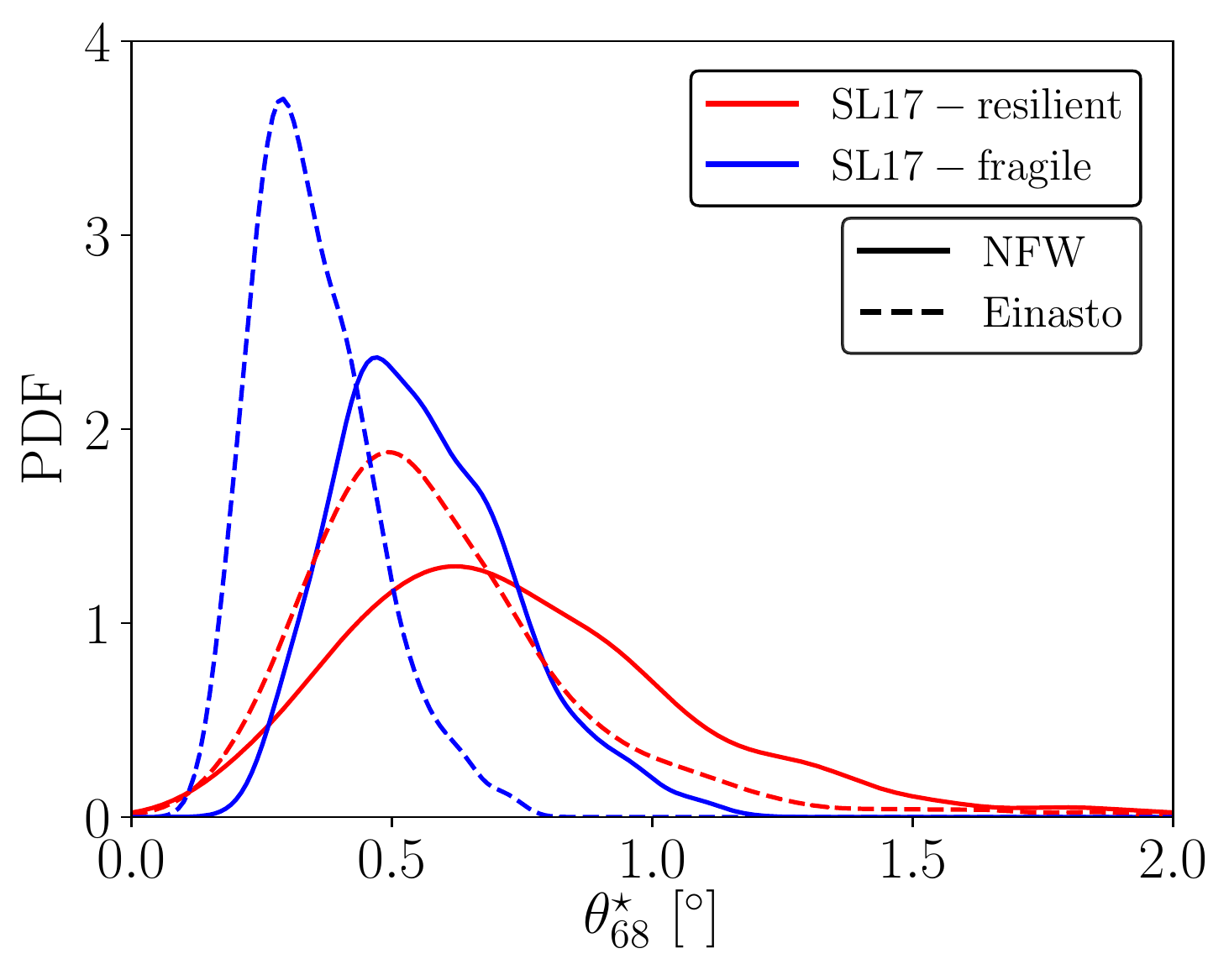}
    \caption{\textbf{Upper panel:} PDF of the total angular size of the brightest subhalo, $\theta_{\rm tot}^\star$. \textbf{Lower panel:} Same as the upper panel for the angle containing 68$\%$ of the total $J$-factor, $\theta_{\rm 68}^\star$.}
    \label{fig:angle_pdf}
\end{figure}

We also compute the PDF of the angular size associated to the brightest subhalos.
The PDF of $\theta_{\rm tot}^\star$, i.e., of the total angular size of the brightest subhalo in each simulation, is shown on the top panel of Fig.~\ref{fig:angle_pdf}. In the SL17-fragile case, the brightest subhalo typically has $\theta_{\rm tot}^\star\sim 2^\circ$ and the PDF is rather narrow, while $\theta_{\rm tot}^\star\sim3^\circ$ for the SL17-resilient model and the PDF is much broader. Note that the choice of profile, NFW or Einasto, does not affect significantly neither the subhalo's radial extension nor its position. The PDF value of $\theta_{\rm tot}^\star$ is thus the same regardless of the density profile shape. On the bottom panel in Fig.~\ref{fig:angle_pdf}, we show the PDF of $\theta_{68}^\star$ which is defined with respect to the radius enclosing 68$\%$ of the total $J$-factor. 
For both subhalo models, the PDF is centered on values smaller than $1^\circ$. 
This is what is expected when computing the radius enclosing 68$\%$ of the total $J$-factor. For an NFW subhalo with tidal radius $r_{\rm t}\gg r_{\rm s}$, we have $\theta_{68}\simeq {\rm arcsin}(r_{\rm s}/(2d))$. For a $10^7\,\rm M_\odot$ subhalo at a distance of $10\,\rm kpc$, this is $\theta_{68}\simeq0.7^\circ$.
For the NFW density profile, $\theta_{68}^\star$ is $0.74^{\circ}$ and $0.56^{\circ}$ for the SL17-resilient and SL17-fragile models, respectively. 
The slightly larger extension of SL17-resilient subhalos compared to SL17-fragile ones is, like their higher $J$-factor, due to their proximity and not to their spatial extension. In fact, the brightest resilient subhalo has in general a smaller tidal radius than the brightest fragile subhalo although the angular extension on the sky is larger.
The central value of  $\theta_{68}^\star$ is slightly smaller for Einasto compared to NFW.

The two subhalo density profiles we consider, NFW and Einasto, are both cuspy. One can wonder what the $J$-factor and angle PDFs would be for subhalos with a cored profile. The SL17 is tailor-made to handle cold DM subhalos as it partly relies on results from cold DM cosmological simulations. Since subhalos have cuspy profiles in these simulations, the model cannot deal with cored subhalos in a consistent way, however we can point out some expected differences. First, a cored subhalo with a given $m_{200}$ and $c_{200}$ is less dense than a cuspy subhalo with the same parameters therefore its $J$-factor is smaller. Second, a lower density also means that cored subhalos are far more susceptible to tidal stripping and disruption, so subhalos in a cored scenario would be less numerous and less extended. 
We therefore leave aside any quantitative estimate for cored subhalo profiles, which would require to run dedicated simulations.

As mentioned already in the introduction, we have no direct information from the simulation
for classifying a subhalo as dwarf galaxy or ``dark satellite".
Nevertheless, we know that, to trigger star formation, a DM subhalo 
should have a mass of around $10^7 - 10^8 \, M_\odot$, depending on the hydrodynamic simulation, see
for example~\cite{Zhu:2015jwa}. If we look at the mass PDF of the brightest subhalo, 
we realize that, in the SL17-fragile model, the brightest halo has a mass typically around
$10^8-10^9 \, M_\odot$, and so it should definitely form a dwarf galaxy. 
On the other hand, in the SL17-resilient model, the mass of the brightest subhalo can be 
lower (down to $10^6 \, M_\odot$), so the halo will not
necessarily form a dwarf galaxy. In this case, because the halo is quite close (closer
than known dwarf galaxies), the $J$-factor can still be very high.
Therefore, whether the brightest halo in the simulation is a dwarf galaxy depends on the
subhalo model (SL17-fragile vs SL17-resilient).
We stress that the nature of the subhalo, being it a dwarf galaxy or optically dark,
does not affect the conclusions reached in the present paper. The possibility
for an extended gamma-ray signal to have a
dwarf galaxy optical counterpart, instead, can contribute 
to firmly identify it as DM subhalo~\cite{Ciuca:2018vsz}.

\section{Simulations of {\it Fermi}-LAT data}
In this section we explain the setup we use to simulate {\it Fermi}-LAT data, the analysis pipeline and the statistical framework that we consider to calculate the significance of the detected signal.

\subsection{Data simulation, background and signal model}
\label{sec:datamock}
We run the full analysis on mock LAT data, realistically simulating background models and the instrument response function, and using state-of-the-art detection pipelines. 

For simulating and analyzing the data, we use {\tt FermiPy}, which is a Python package that automates analyses with the {\tt Fermitools}~\citep{2017ICRC...35..824W}\footnote{See \url{http://fermipy.readthedocs.io/en/latest/}.}.
{\tt FermiPy} is designed to perform several high-level analyses of LAT data such as generating simulations, detecting sources, calculating spectral energy distributions (SED) and finding the source extension.
We employ the {\tt Fermipy} version {\tt 18.0.0} and the {\tt Fermitools} version {\tt 1.1.7}.

We simulate 11 years of gamma-ray data, from 2008 August 4 to 2019 August 4 in the energy range $E=[1,1000]$ GeV.
We consider events belonging to the Pass~8 {\tt SOURCEVETO} event class, and use the corresponding instrument response function {\tt P8R3\_SOURCEVETO\_V2}.
When analyzing the data, we select photons passing standard data quality selection criteria\footnote{\url{https://fermi.gsfc.nasa.gov/ssc/data/analysis/documentation/Cicerone/Cicerone_Data_Exploration/Data_preparation.html}}.
The simulations of gamma-ray data is performed with the {\tt simulate\_roi} tool.
Given a model map (see below for the current model specifications),
this tool takes as input the predicted number of counts for the model, and generates simulated data binned in energy and space. We bin the simulated data with 8 energy bins and angular pixels of size $0.10^{\circ}$.
Using the option {\tt randomize=True} it is possible to randomize the data using Poisson statistics.
We will use {\tt randomize=False}, otherwise differently stated, because we want to test the ideal case of a perfect knowledge of the background components.

We generate mock data sky realizations of a given region of interest (ROI), where we want to test the background-only hypothesis and the background plus signal hypothesis.
We consider two different ROIs, representative of typical background configurations at high Galactic latitudes. We define an ROI of $12^{\circ}\times12^{\circ}$ centered at $(l=150^{\circ},b=60^{\circ})$ for the simulations labelled \textsl{high-latitude}, and at $(l=40^{\circ},b=20^{\circ})$ for the simulations labelled \textsl{low-latitude}.

The astrophysical background model includes the Galactic diffuse emission model, point-like and extended sources selected from the 4FGL catalog~\cite{Fermi-LAT:2019yla}, and an isotropic emission component. 
In particular, we use the Galactic diffuse emission and isotropic templates released, as official ancillary files, with the 4FGL catalog\footnote{\url{https://fermi.gsfc.nasa.gov/ssc/data/access/lat/8yr_catalog/}}.

The signal model is represented by a DM subhalo,  centered at the center of the ROI (either \textsl{high-} or  \textsl{low-latitude}). 
The spectrum of the DM injected signal is normalized by the thermally averaged annihilation cross section, $\langle \sigma v \rangle$, and depends on the mass of the DM particle (we test masses of 10, 100, 1000 GeV). 
We use a benchmark annihilation channel into $b$-quarks.\footnote{ \url{https://fermi.gsfc.nasa.gov/ssc/data/analysis/scitools/source_models.html\#DMFitFunction}}
We vary the value of the annihilation cross section from 10$^{-27}$ up to 10$^{-22}$ cm$^3$/s, to check how the detection sensitivity changes with the brightness of the signal.
The spatial distribution of the DM signal is  
 built from Eq.~(\ref{eq:J}). In order to get an estimate of the uncertainties at play, we select one hundred subhalos within 1$\sigma$ of the mean of the $J$-factor PDF shown in Fig.~\ref{fig:corr_jfact_pdf}.
The analysis is then repeated by using the spatial template corresponding to each of these $J$-factors, as injected signal.
We consider four different models varying the impact of tidal disruption (SL17-fragile or SL17-resilient) and the subhalo density profile (NFW or Einasto). 

If not otherwise specified, we adopt as baseline configuration for signal injection the SL17-resilient subhalo model, an NFW DM subhalo density, a DM mass of 100 GeV, and the \textsl{high-latitude} ROI. 

\subsection{Signal reconstruction models}
\label{sec:pipeline}
We perform a fit on simulated data using the {\tt gta.fit} tool, which is a wrapper of the {\tt pyLikelihood} fit method implemented in the {\tt Fermitools}.
This tool returns the best fit and error of SED parameters and the full covariance matrix of the model.
From the fit, we extract the value of the log-likelihood to estimate the significance of the detection of the DM signal with respect to the background-only hypothesis.

In order to reconstruct the injected DM signal and study the detectability of DM subhalos, we make different (spectral and spatial) assumptions on the reconstructed signal.
Since, in real data analyses we cannot know the distance and density profile parameters of the subhalo, we try first to detect it as a point-like source (\texttt{PS}). The fitted SED is a power law, and the free model parameters are normalization and spectral index. 
Secondly, we test the hypothesis of an extended source (\texttt{Ext}) for which we use a radial Gaussian as spatial template, and a power-law SED. In this case, the free parameters are the same as the \texttt{PS} case with, in addition, the size of extension, namely the width of the radial Gaussian spatial model.
For the extended models, the source extension is computed using the {\tt gta.extension} tool that performs fits to the data with different sizes of Gaussian and then maximizes the log-likelihood as a function of this parameter, by leaving the SED parameters free as well. 

In addition, we test a case in which we have a point-like source plus an extended source, centered at the same position (\texttt{PS+Ext}). Both components have SED modeled as power laws\footnote{A priori the SED of the two components should be constrained to be unique, in order to claim that this is a single source. This being of impractical implementation in {\tt FermiPy}, we checked a posteriori that, indeed, the best-fit spectral indices are compatible within 1$\sigma$. We also notice that the best-fit SED of the \texttt{PS} and \texttt{Ext} models are compatible.}. In this case, the free parameters are normalizations, spectral indices and sizes of extension of the two components.
We summarize models and parameters in Tab.~\ref{tab:modelfits}. 
For all the model tested we do not vary the position of the source.

\begin{figure}
\includegraphics[width=0.49\textwidth]{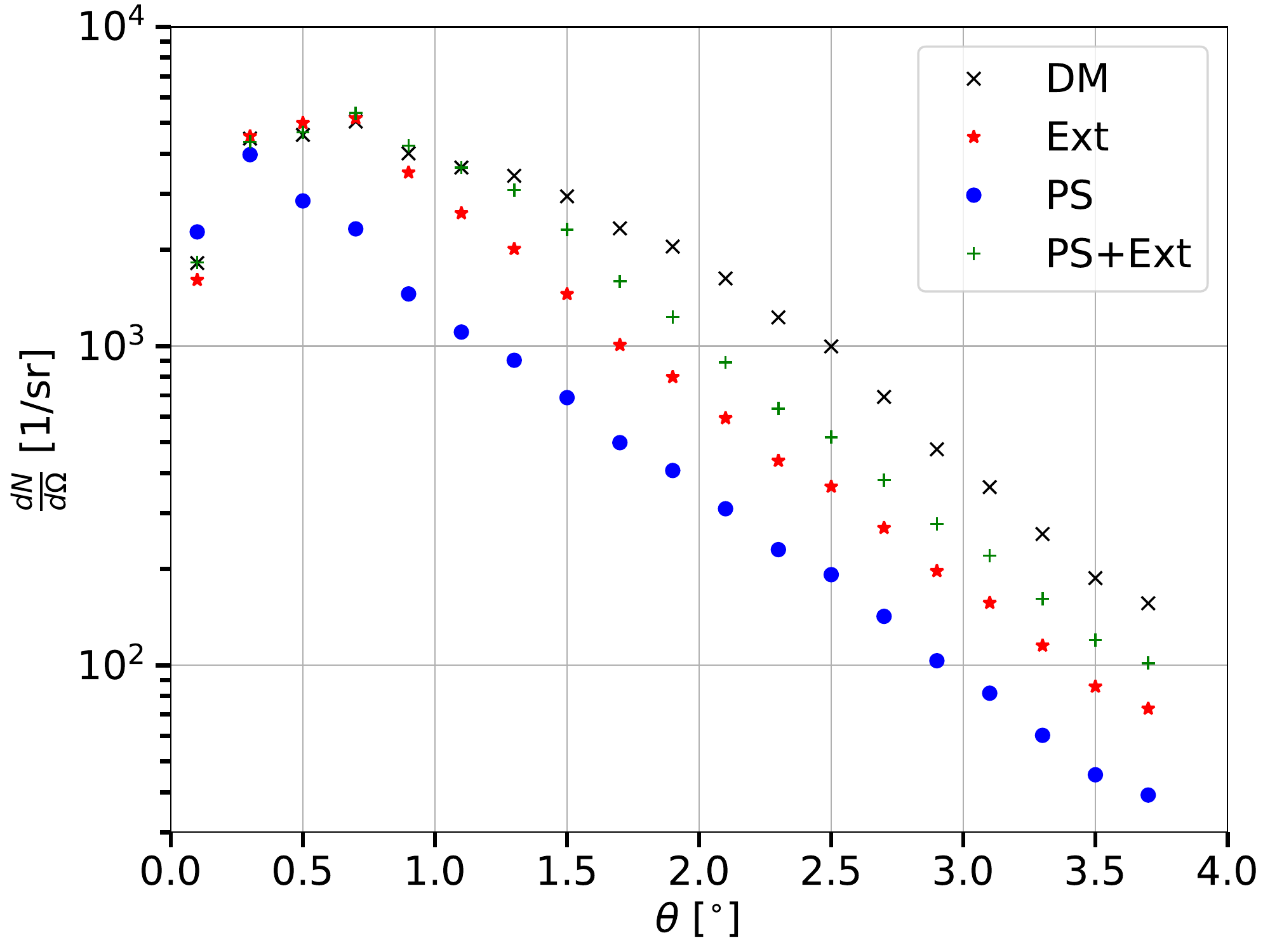}
\caption{Surface brightness angular profile of the injected DM signal (black crosses), compared with the angular profiles of the best-fit \texttt{PS} (blue circles), \texttt{Ext} (red stars), and  \texttt{PS+Ext} (green plus signs) models.
}  
\label{fig:SB}
\end{figure}

In Fig.~\ref{fig:SB}, we compare the angular profiles, calculated as the surface brightness (number of counts per solid angle as a function of the angular distance from the ROI center), of the three models adopted for the signal reconstruction, together with the angular profile of the DM injected signal.
For each source template, we compute the number of counts for concentric annuli in angular distance and we divide the number of counts in each annulus by its solid angle. We obtain with this method the surface brightness of the signal. All templates have been convoluted with the instrument angular resolution (point spread function) and are normalized such as to match the DM injected signal at the peak.
The \texttt{PS} case provides the poorest fit to the DM emission. In fact, this model produces a flux comparable with the injected signal only in the inner $0.3^{\circ}$, instead at higher angular distances the surface brightness is much smaller than the DM signal. Instead, the \texttt{Ext} starts to deviate significantly from the DM injected signal at distances $>1^{\circ}$ and also in the inner $0.2^{\circ}$ it slightly underestimates the flux. On the other hand, the  \texttt{PS+Ext} case fits well the DM signal up to $\sim 1.3^{\circ}$. Also, it well matches the DM injected signal in the inner $0.5^{\circ}$. 
A point source plus and extended source can therefore provide the best fitting model of the DM injected signal. 
We will test this possibility in the next section.

\begin{table}
\caption{Models used for the reconstruction of the DM signal: Model name, SED parameterization, spatial distribution (morphology), and number of free parameters in the fit.}
\begin{tabular}{ |c|c|c|c| } 
 \hline
 Model & Spectrum & Morphology &  No.~params \\ 
 \hline
 \hline
 \texttt{PS} & power-law & point-like & 2 \\ 
 \hline
 \texttt{Ext} & power-law & radial Gaussian  & 3\\ 
 \hline
 \texttt{PS+Ext} & power-law &  point-like + radial Gaussian  & 5\\ 
 \hline
\end{tabular}
\label{tab:modelfits}
\end{table}

\subsection{Statistical significance}
\label{sec:sign}
Our null hypothesis ($H_0$) is defined by the background-only model, when we fit the simulated data without any additional DM signal.
The alternative hypothesis is instead represented by our reconstructed signal templates (as described above), through which we test the presence of an additional source on top of the astrophysical background.

The reconstructed signal models are all nested models for which Wilk's theorem~\cite{wilks1938} usually applies. As usually done, we define the Test Statistics as $TS = 2 \, (\log_{10} \mathcal{L}_{H_0}- \log_{10} \mathcal{L}_{H_1})$.
However, Wilk's theorem cannot be applied if some of the parameters of the test hypothesis (in the limit of the null hypothesis) take values on the boundary of the allowed parameter space.
As an example, for the \texttt{PS} case, the free fit parameters are the normalization and the spectral index of the SED. This model reduces to the null hypothesis when the normalization tends to zero, which corresponds to the lower bound of its permitted value. 
If this is the case, the $TS$ distribution is given by a mixed distribution, which depends on the number of parameters whose null value is restricted to be at the boundary of the allowed range, and on those which are not.
Following \cite{10.2307/2289471,Macias:2016nev} we can calculate the $p$-value for a given $TS$ as:
\begin{equation}
p(TS) = 2^{-n} \left( \delta(TS) + \sum_{i = 1}^{n} \binom{n}{i} \chi^2_{i +k}(TS) \right) \, ,
\label{eq:sig}
\end{equation}
where $n$ is the number of restricted parameters (i.e. the parameters that have a boundary condition in the limit of the null hypothesis) and $k$ is the number of unrestricted parameters.

We report in Tab.~\ref{tab:nkparams}, the number of restricted and unrestricted parameters for the hypotheses we compare one with respect to the other.

\begin{table}
\caption{Number of restricted and unrestricted ($n; k$) parameters used in Eq.~\ref{eq:sig}, for the calculation of the statistical significance}.
\begin{tabular}{ |c|c|c|c| } 
 \hline
  &   \texttt{PS}  &  \texttt{Ext}  &  \texttt{PS+Ext} \\ 
 \hline
 $H_0$ & (1; 1) & (1; 2) & (2; 3) \\
 \hline
 \texttt{PS} & -- & (1; 1) & (1; 2) \\ 
 \hline
 \texttt{Ext} & --   & -- & (1; 1)\\ 
 \hline
\end{tabular}
\label{tab:nkparams}
\end{table}

\section{Results}
\label{sec:results}

\begin{figure}
\includegraphics[width=0.49\textwidth]{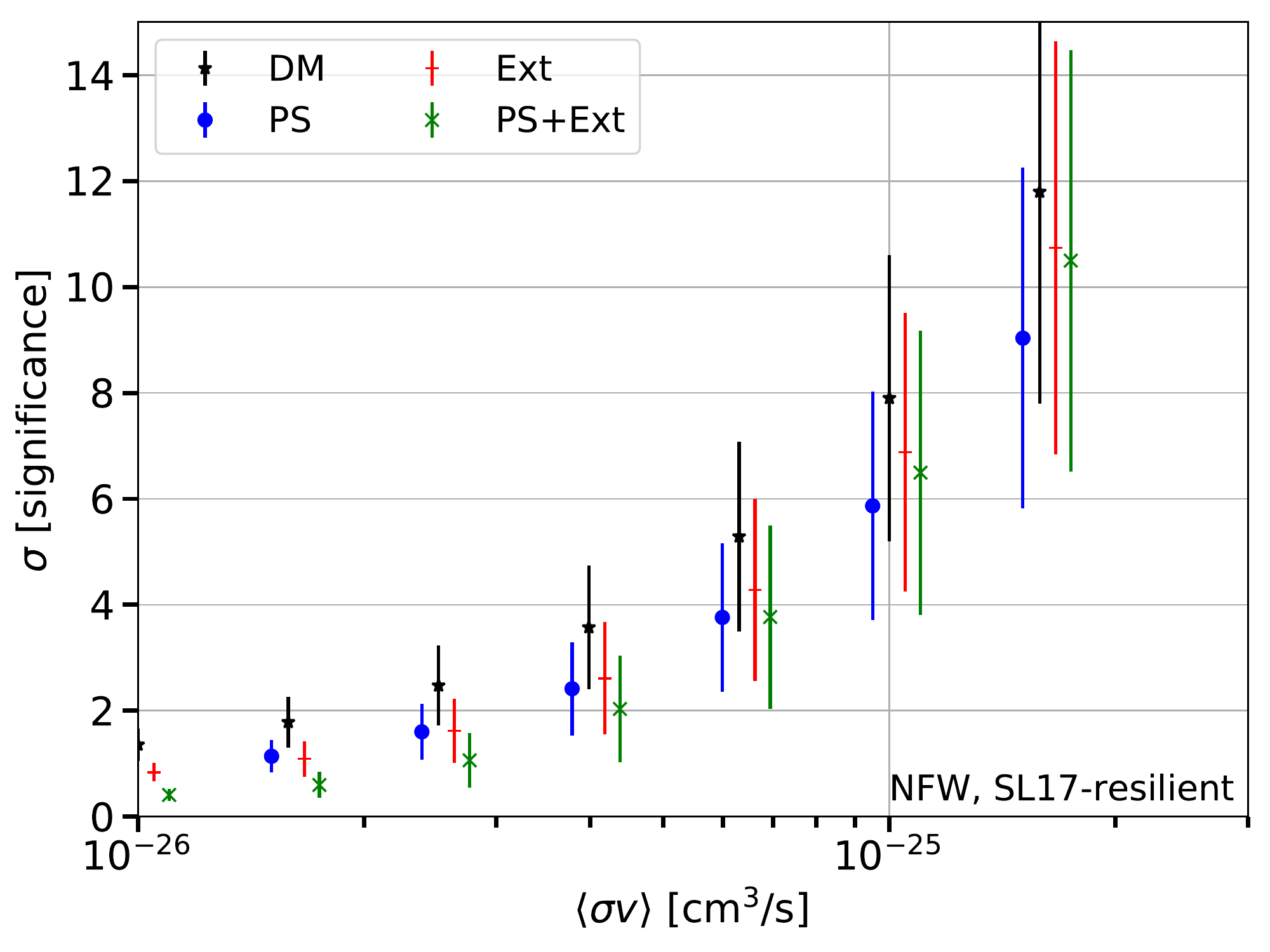}
\caption{Detection significance as a function of the injected signal annihilation cross section, for different signal reconstruction models (\texttt{PS} in blue, \texttt{Ext} in red, \texttt{PS+Ext} in green, \texttt{DM} in black). The cross sections used for signal injection correspond to the abscissas of the black points; for the other cases the shift along the $x$ axis is for visual ease only (this is true for all other plots in the paper). 
The subhalo model adopted is SL17-resilient, and we assume an NFW density profile for the subhalos. }  
\label{fig:cases}
\end{figure}

As described in Sec.~\ref{sec:datamock}, we create mock data making different assumptions on the DM injected signal, besides testing different source models in the fit.
In Fig.~\ref{fig:cases}, we show the detection significance (significance of the source model with respect to $H_0$) for the different signal reconstruction templates (\texttt{PS}, \texttt{Ext}, \texttt{PS+Ext}), and as a function of the injected signal cross section. 
For illustrative purposes only, we add the case \texttt{DM}, in which we fit mock data with the same DM template used to simulate them, leaving free only the overall normalization of the signal (one restricted parameter).
The subhalo model adopted here is SL17-resilient, and we assume an NFW density profile for the subhalos. 
From this figure we can see at first that, even in the optimistic (as well as unrealistic) case in which we know everything about subhalo properties and position (\texttt{DM}), we could reach a detection significance larger that 3$\sigma$ (marginal hint) for annihilation cross sections above $3 \times 10^{-26}$ cm$^3$/s. 
Such cross sections (for annihilation into $b$-quarks and DM mass of 100 GeV, i.e.~our reference case) are still allowed 
by current constraints coming from the observation of dwarf spheroidal galaxies~\cite{Fermi-LAT:2016uux, Calore:2018sdx}, as well as of the Galactic halo at high latitudes~\cite{Zechlin:2017uzo,Chang:2018bpt}. 
Moreover, given the similar results we get for point-like and extended templates, such a sensitivity estimate is compatible with what is found for the \Fermi-LAT sensitivity to point-like DM subhalos, e.g.~\cite{Calore:2016ogv,Calore:2019lks}.
A firm detection (above 5$\sigma$, without accounting for look-elsewhere effects) would instead need cross sections at least as high as $5-6 \times 10^{-26}$ cm$^3$/s -- which, again, is not excluded by current gamma-ray constraints.
Below the 3$\sigma$  \texttt{DM}  detection significance threshold, all models provide comparable evidence for DM subhalos, as expected.
For low cross sections the log-likelihood for the \texttt{PS} and \texttt{Ext} cases are  very similar. Since the \texttt{PS} template has less parameters, it gives a slightly higher detection significance.
Above cross sections of $3 \times 10^{-26}$ cm$^3$/s instead the extended template, \texttt{Ext}, starts to provide the best fit among the three reconstructed signal models, with the \texttt{PS+Ext} model giving comparable detection significance.

If not stated otherwise, in what follows, we present results for the \texttt{Ext} template.
The \texttt{PS+Ext} case would produce very similar results -- with $< 2\sigma$ improvement of the fit when adding a point source component to the extended source for cross sections below $2 \times 10^{-25}$ cm$^3$/s.

\begin{figure}
\includegraphics[width=0.49\textwidth]{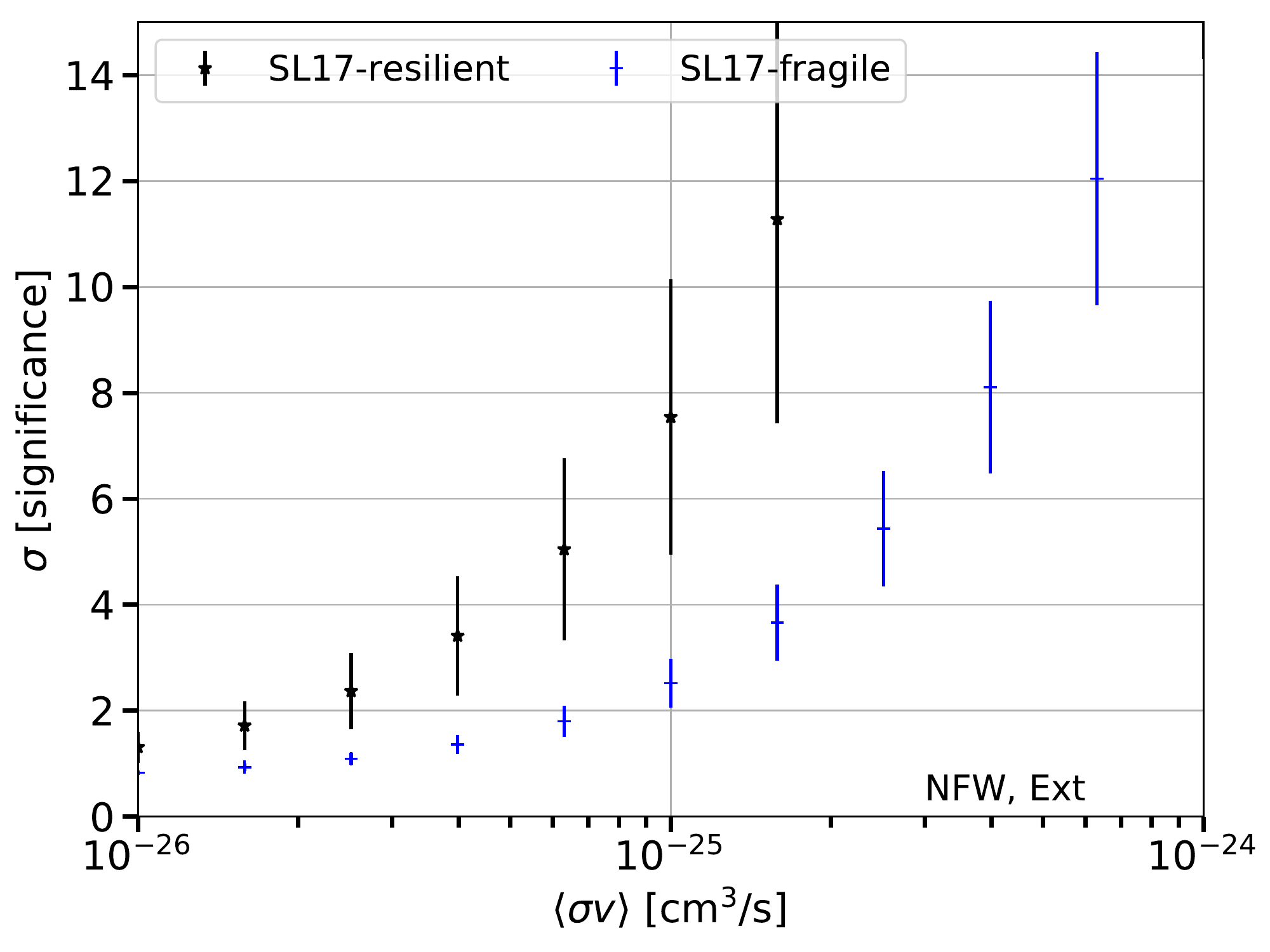}
\caption{Detection significance for the \texttt{Ext} signal reconstruction model comparing SL17-resilient (black) and SL17-fragile (blue) subhalo models.}  
\label{fig:fragres}
\end{figure}

As presented in Sec.~\ref{sec:sim} the SL17-resilient and SL17-fragile models bracket the uncertainty in the modeling of tidal disruption of Galactic DM subhalos.
In Fig.~\ref{fig:fragres} we compare the detection significance obtained with the two subhalo models (for an NFW DM subhalo density profile). The SL17-resilient model provides a much higher detection significance regardless of the injected signal cross section. The difference in significance at fixed cross section is roughly a factor of $\sim 3$. Indeed, for a cross section of $4\times 10^{-26}$ cm$^3$/s the SL17-fragile model gives $1\sigma$ detection significance while the SL17-resilient almost $4\sigma$.
This difference in detection significance can 
be understood by looking at the difference in the $J$-factor distribution, cf.~Fig.~\ref{fig:corr_jfact_pdf} (bottom panel), and can have an impact in the interpretation of the results of real data analyses. 
In light of this result, detecting subhalos with \Fermi-LAT in the SL17-fragile scenario, while respecting the constraints from other targets, seems quite challenging.

\begin{figure}
\includegraphics[width=0.49\textwidth]{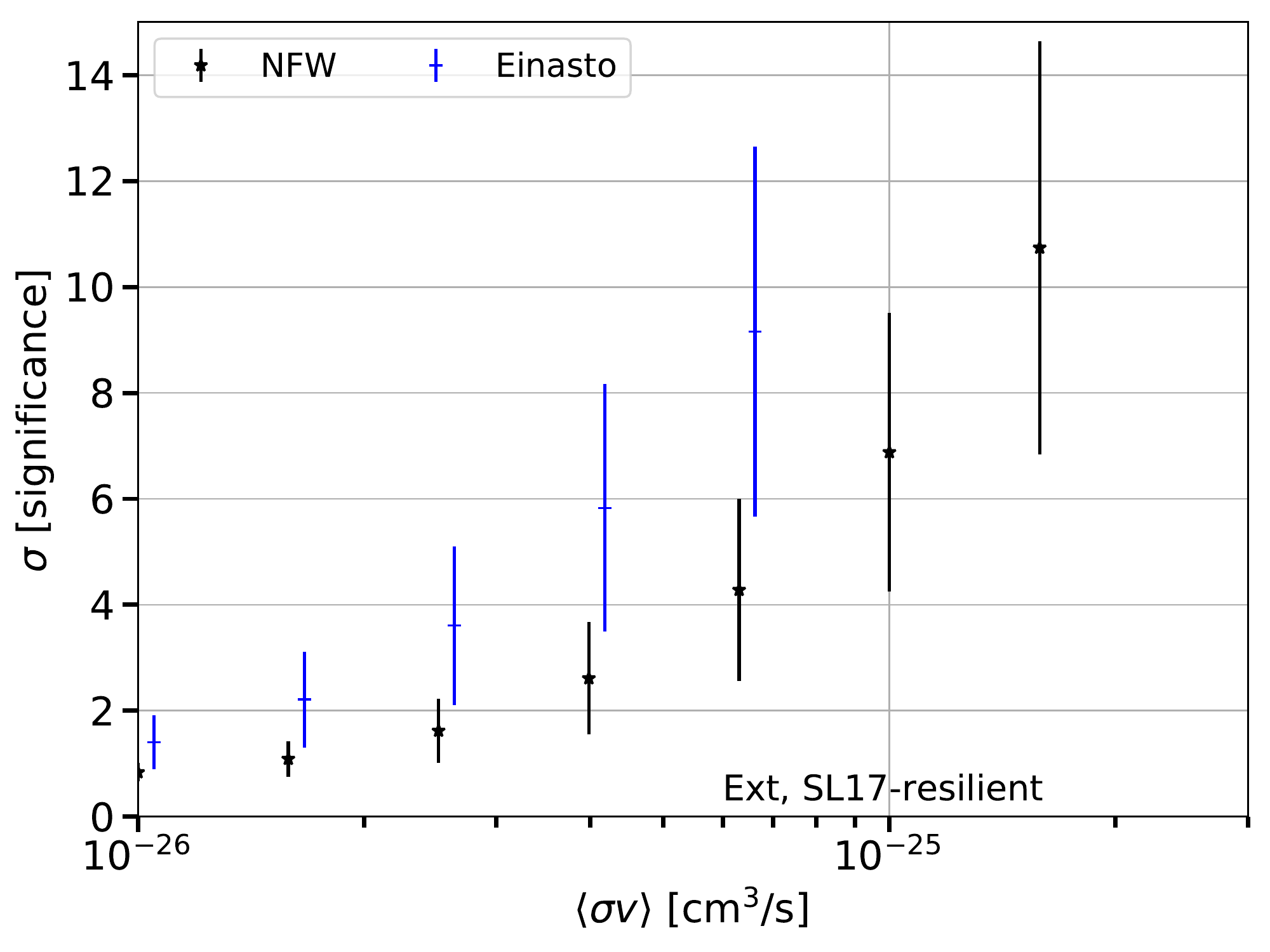}
\caption{Detection significance for the \texttt{Ext} signal reconstruction model comparing NFW (black) and Einasto (blue) subhalo density profiles, for the SL17-resilient subhalo model. }
\label{fig:einnfw}
\end{figure}

Fig.~\ref{fig:einnfw} shows the comparison between NFW and Einasto subhalo density profiles, for the SL17-resilient subhalo model.
For a given cross section, the detection significance obtained with an Einasto profile is always larger than the one found with the NFW, by roughly a factor of 2, cf.~Fig.~\ref{fig:corr_jfact_pdf} (bottom panel).

\begin{figure}
\includegraphics[width=0.49\textwidth]{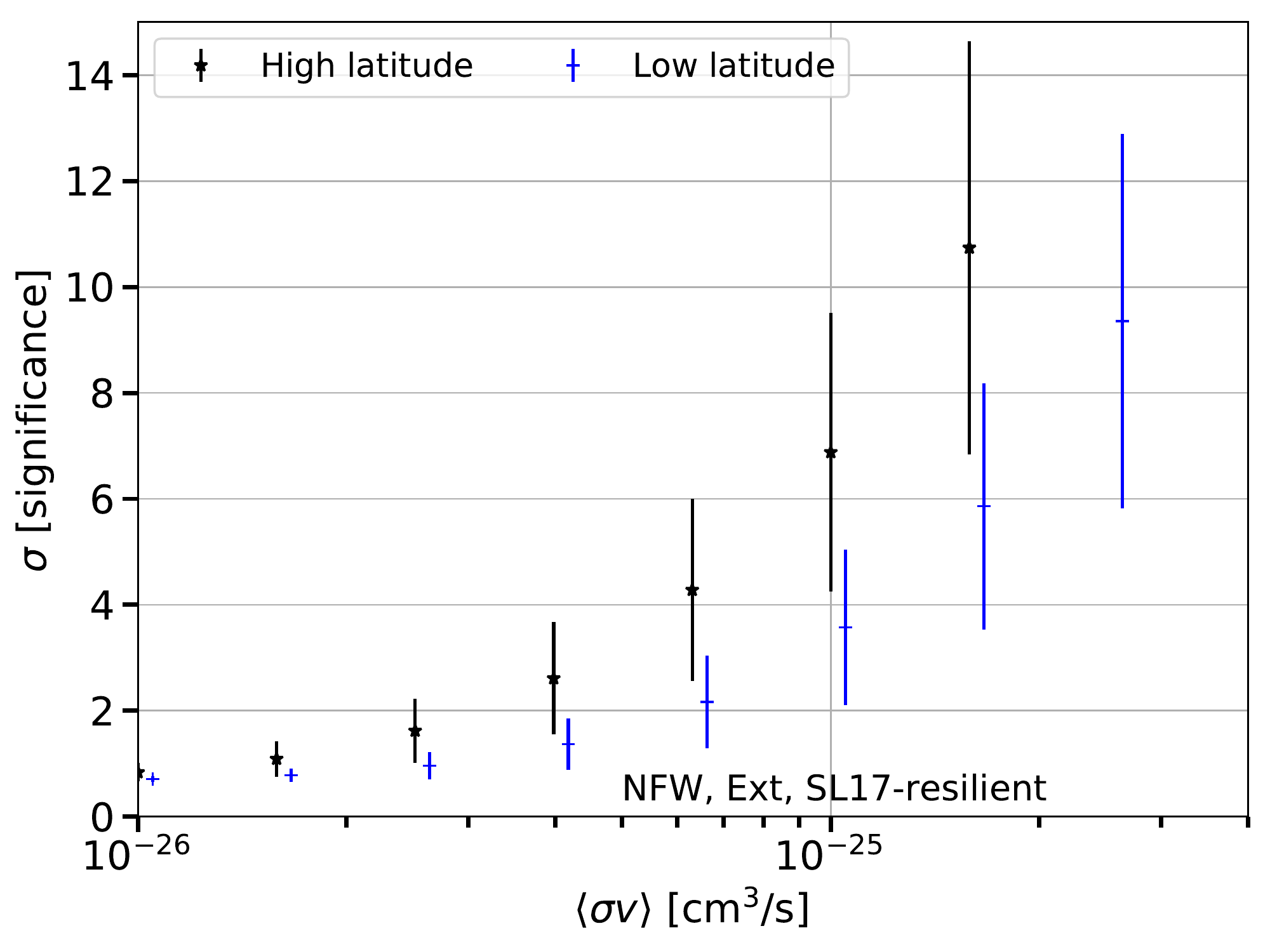}
\caption{Detection significance for the \texttt{Ext} signal reconstruction model comparing subhalos located at two different positions in the Galaxy (see the text for further details).}  
\label{fig:ROI}
\end{figure}

\begin{figure}
\includegraphics[width=0.49\textwidth]{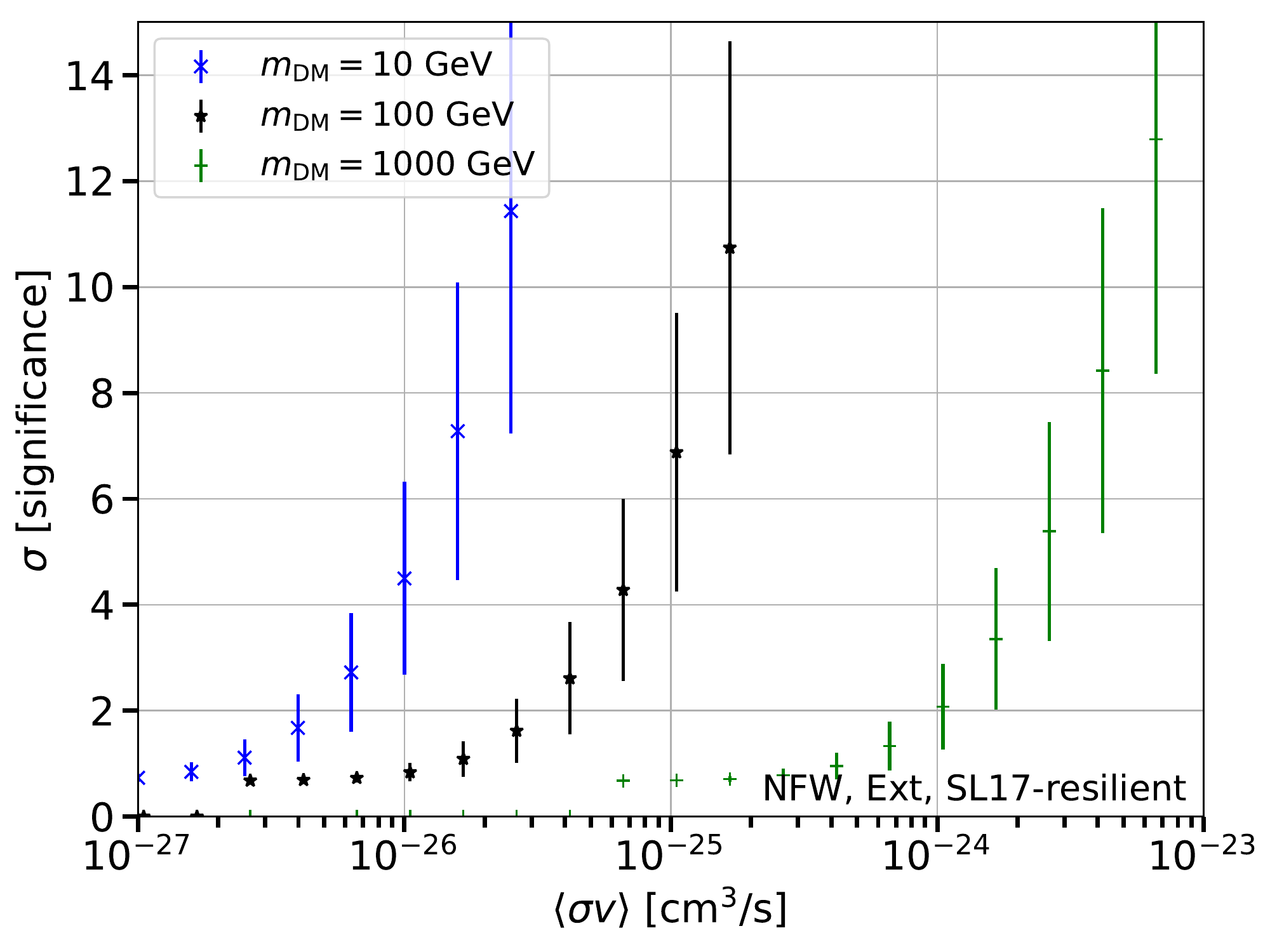}
\caption{Detection significance for the \texttt{Ext} signal reconstruction model comparing different DM masses for the injected signal.}  
\label{fig:mass}
\end{figure}

Finally, we show the results obtained by placing the DM subhalo at the center of the \textsl{low-latitude} ROI, cf.~Fig.~\ref{fig:ROI}. 
As expected, it is much easier to detect a subhalo (even if extended) at high latitudes than at lower latitudes, where the background from interstellar emission is more intense. The DM subhalo should therefore have larger $\langle \sigma v \rangle$ to produce the same significance as the \textsl{high-latitude} ROI case.

In Fig.~\ref{fig:mass}, we instead compare the detection sensitivity for different choices of the DM mass. 
In this case, considering different masses shifts the results along the $\langle \sigma v \rangle$ values. In particular, for a fixed cross section the lower is the mass the higher is the significance for the detection of a subhalo. 
This is explained by the fact that a less (more) massive DM with respect to the benchmark case (100 GeV) produce a gamma-ray spectrum with a peak at lower (higher) energies.
{\it Fermi}-LAT has a peak of the sensitivity at about 2-4 GeV. Instead at higher energies the sensitivity increases monotonically\footnote{See this page for the description of the LAT sensitivity as a function of energy \url{https://www.slac.stanford.edu/exp/glast/groups/canda/lat_Performance.htm}}. Therefore, DM candidates with a peak of the spectrum at a few GeV, such as $b\bar{b}$ annihilation channel with $m_{\rm{DM}}=10$ GeV, are detected with the highest significance while candidates with the peak at higher energies have low significance values.

\begin{figure}
\includegraphics[width=0.49\textwidth]{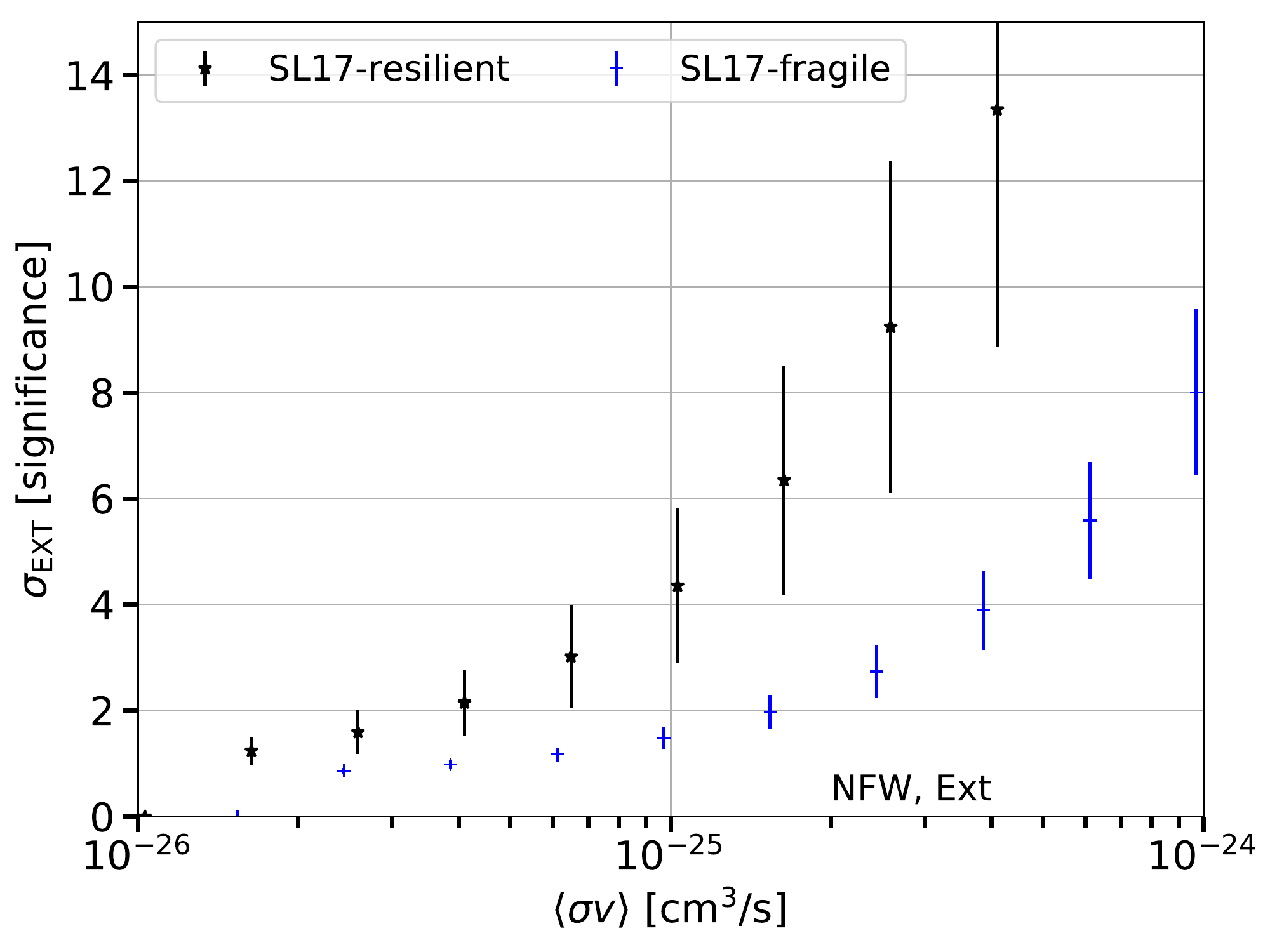}
\includegraphics[width=0.49\textwidth]{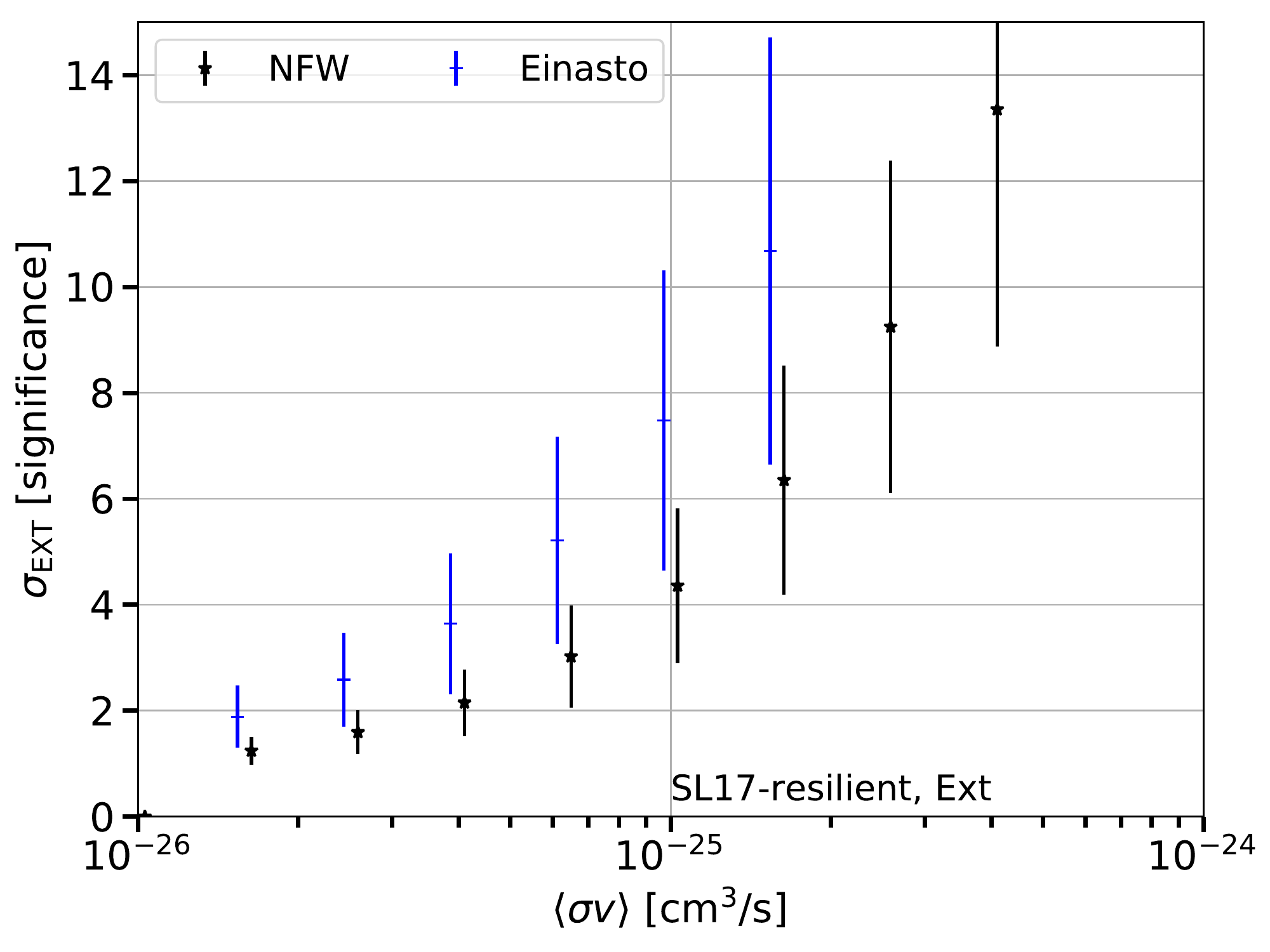}
\caption{Significance of extension ($\sigma_{EXT}$) of the \texttt{Ext} template w.r.to the \texttt{PS} one, for the subhalos of our simulations. {\bf Top panel}: Comparison between the SL17-resilient (black) and SL17-fragile (blue) subhalo models, for an NFW DM subhalo profile.  {\bf Bottom panel}: Comparison between an Einasto (blue) and NFW (black) DM subhalo profile, for the SL17-resilient subhalo model.}  
\label{fig:signext}
\end{figure}

\begin{figure}
\includegraphics[width=0.49\textwidth]{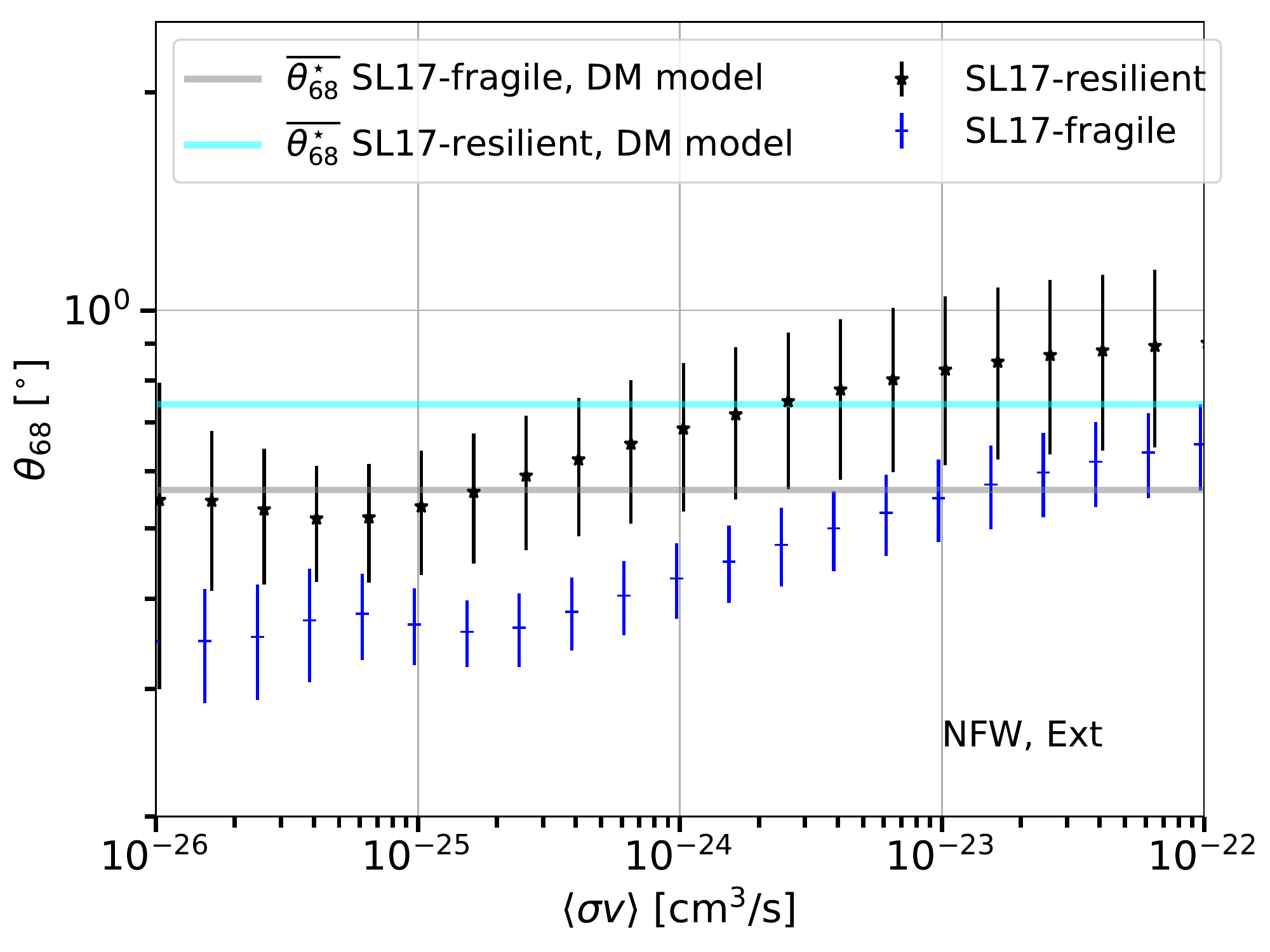}
\includegraphics[width=0.49\textwidth]{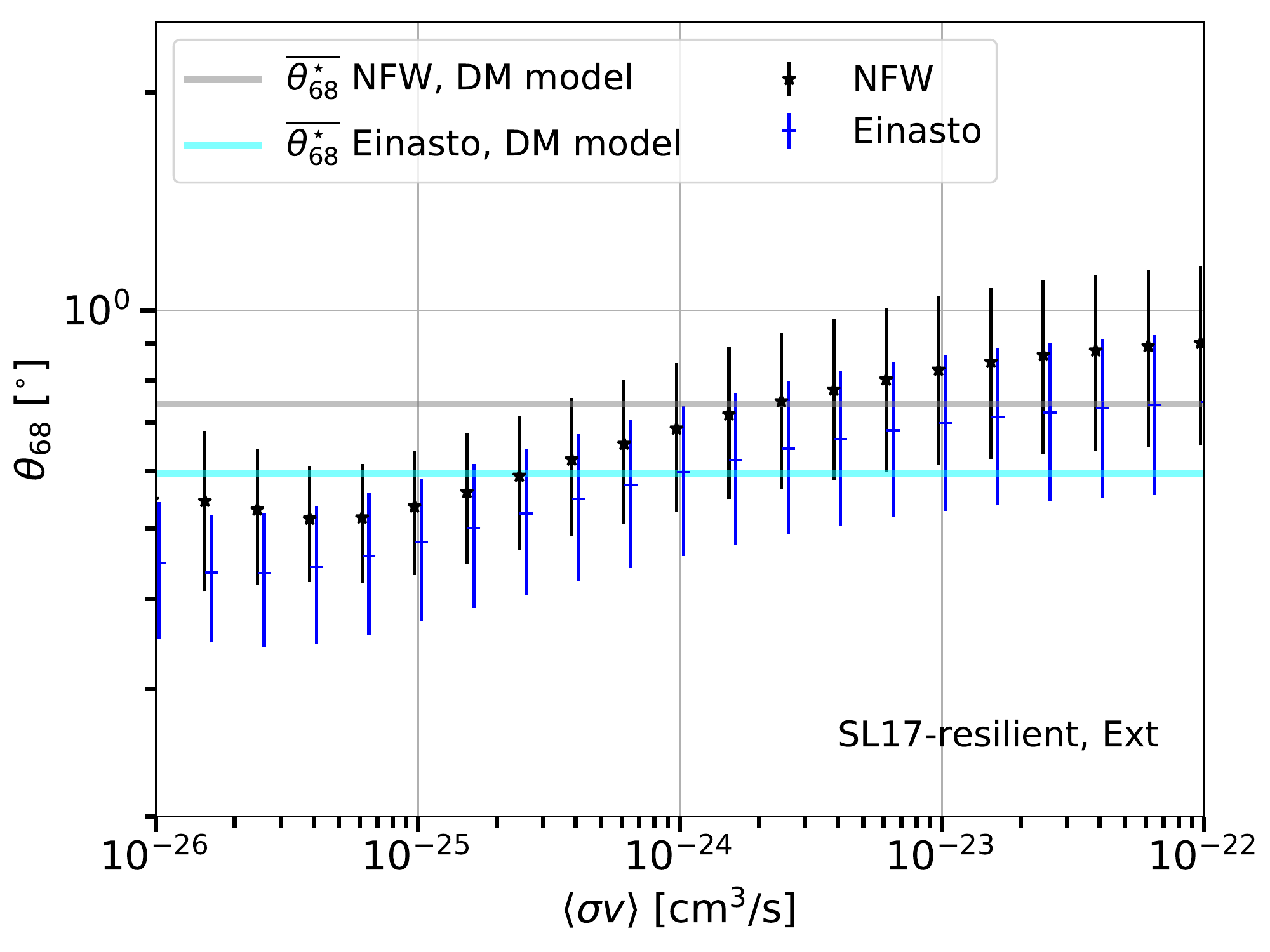}
\caption{\textbf{Top panel:} Reconstructed $68\%$ containment radius ($\theta_{68}$) as a function of the injected annihilation cross section, comparing the SL17-resilient (black) and SL17-fragile (blue) models for an NFW subhalo density profile.
\textbf{Bottom panel:} Same as the top panel, but for the SL17-resilient model comparing NFW (black) and Einasto (blue) DM density profiles.
In both panels, we also overlay theoretical predictions for the average value of $\theta_{68}^\star$, i.e.~$\overline{\theta_{68}^\star}$.}  
\label{fig:ext}
\end{figure}

Up to this point we have demonstrated that the brightest DM subhalo can be detected with the highest significance when fitted with an extended source template, either \texttt{Ext} or \texttt{PS+Ext}.
Nevertheless, a legitimate question to ask is: Is the evidence for extension significant? 

We quantify the significance for the extension of our simulated signal. We consider the case of one extended source (\texttt{Ext}) fitted to the DM signal. Similar results are found if we consider the case with one point source and one extended source (\texttt{PS+Ext}).

In Fig.~\ref{fig:signext} we show the significance for the extension of the source ($\sigma_{\rm{EXT}}$). This is calculated by considering the \texttt{PS} case as null hypothesis and the extended source case, \texttt{Ext}, as test hypothesis. We compute the significance for extension following the procedure highlighted in Sec.~\ref{sec:sign}.
We vary several assumptions on the simulated DM signal model: We compare SL17-resilient and SL17-fragile subhalo models (top panel), and the choices of different DM subhalo density profiles for the SL17-resilient case (bottom panel).
The SL17-fragile model provides much lower $\sigma_{\rm{EXT}}$ with respect to the SL17-resilient case. For example, at a cross section of $10^{-25}$ cm$^3$/s the SL17-fragile subhalo model gives an average significance for extension of 1.8$\sigma$ while SL17-resilient gives 4.2$\sigma$. 
On the other hand, the results obtained for Einasto and NFW profiles are comparable. Indeed, for $\langle \sigma v \rangle = 10^{-25}$ cm$^3$/s we get, on average, a significance for extension of 7.5$\sigma$ and 4.2$\sigma$, respectively.
In all cases, but the SL17-fragile, a marginal detection for extension ($\sim 3\sigma$) is achieved for cross sections 
 $3-4 \times 10^{-26}$ cm$^3$/s, which are values still permitted by current constraints, as seen above. 
 For the SL17-fragile case a $3\sigma$ detection of extension requires, instead, cross sections of about $3 \times 10^{-25}$ cm$^3$/s, which starts to be in tension with current constraints from dwarf spheroidal galaxies.

In Fig.~\ref{fig:ext} we show the reconstructed $68\%$ containment radius ($\theta_{68}$, also equivalent to the standard deviation of the radial Gaussian template) for different hypotheses on the
injected DM signal.
In the same plot, we also overlay the theoretical values $\overline{\theta_{68}^\star}$ corresponding to the mean over the $\theta_{68}^\star$ distribution of the sampled halos, cf.~Sec.~\ref{sec:sim}.
We note that the reconstructed $\theta_{68}$ increases with the brightness of the injected signal until it reaches a plateau, which is compatible (within the $1\sigma$ error band) with the theoretically predicted value.\footnote{We stress however that the theoretically predicted value is computed without convolving the DM template with the point spread function of the instrument, and so it is expected that the measured $\theta_{68}$ is slightly larger than the predicted $\theta_{68}^\star$.}
Indeed, if the DM subhalo signal is too faint the analysis picks up only the more central part of the emission and thus the size of extension is lower than the simulated one. This trend is visible for all the cases considered in this analysis implying that for a faint DM signal the size of extension is underestimated. 
We stress, however, that, as shown above, even in the case of faint signals where the extension may be underestimated the evidence for the extension is significant -- namely above $3\sigma$ for cross sections above $3 \times 10^{-26}$ cm$^3$/s in the SL17-resilient case.
At the plateau, the size of extension is roughly $0.8^{\circ}$ for the SL17-resilient and $0.6^{\circ}$ for the SL17-fragile cases and NFW profile, while the theoretical values are $0.74^{\circ}$ and $0.56^{\circ}$, respectively. The Einasto and NFW density profiles give very similar results with the Einasto profile which produces slightly smaller values for $\theta_{68}$ ($0.59^{\circ}$ for the SL17-resilient subhalo model).

\begin{figure}
\includegraphics[width=0.49\textwidth]{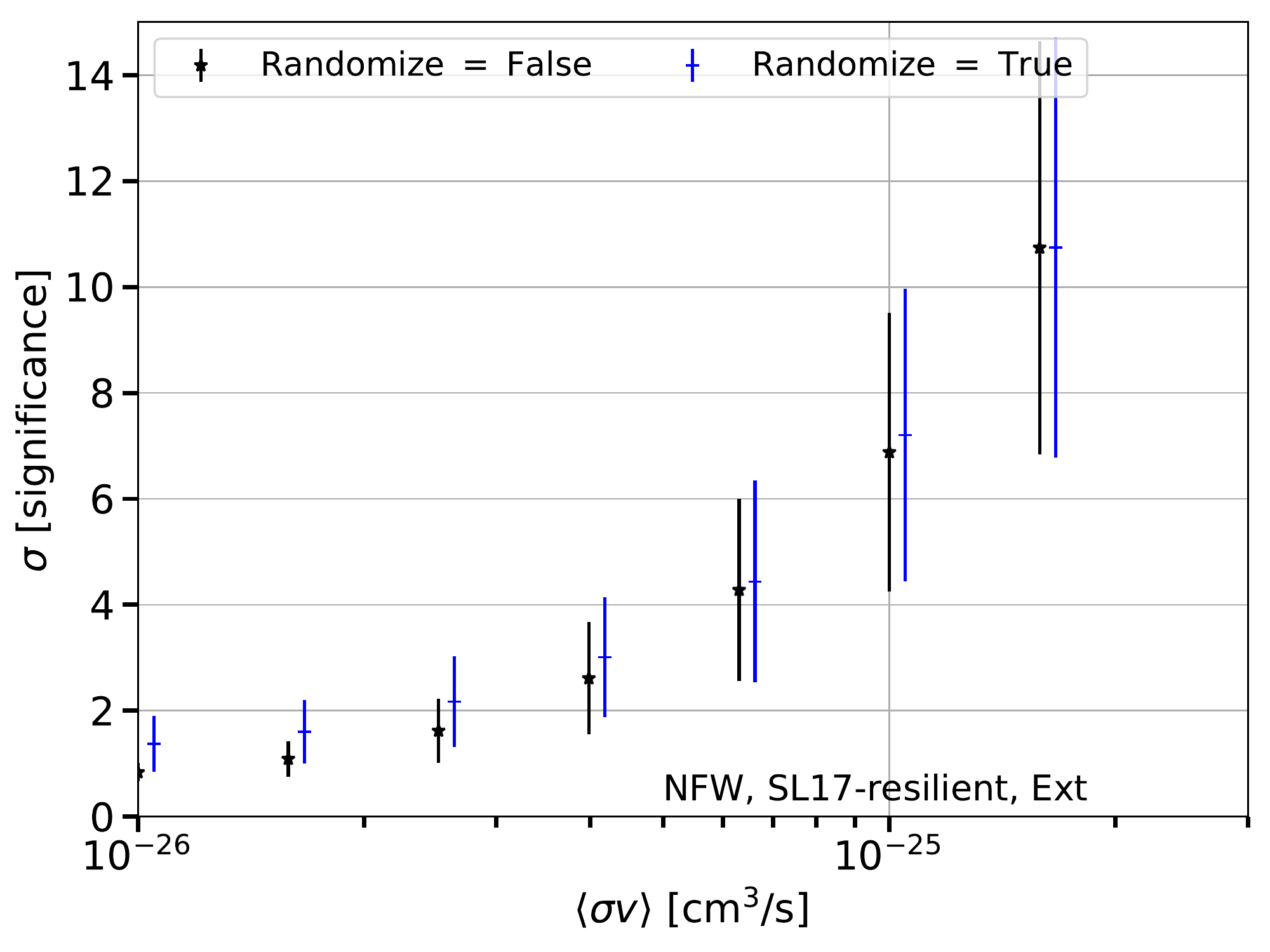}
\caption{Same as Fig.~\ref{fig:cases} for the case in which we randomize (blue) or not (black) the number of counts in each pixel, according to Poisson statistics.}  
\label{fig:random}
\end{figure}

Finally, we study how the signal reconstruction is affected by randomizing simulated data counts using Poisson statistics, i.e.~\texttt{randomize=True}. The result is shown in Fig.~\ref{fig:random} for the \texttt{Ext} case. We see that the detection significance is not affected by randomization and, therefore, all conclusions reached above still hold in the case of added random Poisson noise.

\section{Discussion and conclusions}
\label{sec:conclusions}
A general ``belief" is that among {\it Fermi}-LAT unidentified sources may shine DM subhalos, although the majority of those should be active galactic nuclei or other galaxies that lack, at the moment, detection in other wavelengths. With the present work, we re-assessed the sensitivity of the LAT to signals from the brightest DM subhalo, in the light of the fact that subhalos with the highest $J$-factor show a significant extension in the sky -- as supported by a correlation between subhalo angular extension and $J$-factor. 

We quantified the sensitivity of {\it Fermi}-LAT to the brightest {\it extended} DM subhalo, by performing realistic simulations of the DM injected signal and analysis reconstruction.
We tested different assumptions for the DM subhalo model (SL17-resilient and SL17-fragile) and density profile (NFW and Einasto), as well as different DM masses for the DM injected signal. 
We fit the DM subhalo source with three different signal reconstruction templates: \texttt{PS}, \texttt{Ext} and \texttt{PS+Ext}.

Our results show that:
\begin{itemize}
    \item For both the SL17-resilient and SL17-fragile models, above 3$\sigma$ detection significance the extended template, \texttt{Ext}, always provides the best fit among the three reconstructed signal models, and also gives a detection significance comparable to the one we would get by fitting the DM injected  signal  with a perfectly known DM template.  
    A firm detection (above 5$\sigma$, without accounting for look elsewhere effects) of DM \textsl{extended} subhalos for the SL17-resilient model can be made for cross sections at least as high as $5-6 \times 10^{-26}$ cm$^3$/s (100 GeV DM mass), which are not excluded by other gamma-ray constraints yet.
    \item  The values of the annihilation cross section increase by about a factor of four if we consider, instead, the SL17-fragile subhalo model. This implies that accounting for uncertainty on the subhalo model is a crucial step towards a correct interpretation of DM searches in real data, and that the detection of extended subhalos in the SL17-fragile scenario would be challenging, while fulfilling other gamma-ray bounds on the annihilation cross section. On the other hand, the results are not very sensitive to changing the DM density profile within subhalos. In particular using an Einasto or NFW profile provides compatible detection significance. 
    \item  The evidence for extension is \textsl{always significant} for cross sections above $3 \times 10^{-26}$ cm$^3$/s (SL17-resilient case, NFW and Einasto profiles). In particular, the reconstructed extension for bright signals is compatible with the theoretical expectation from subhalo simulations, while it is slightly underestimated for faint signals.
  \end{itemize}
    
    In the most optimistic case, we showed that for cross sections still allowed by other gamma-ray constraints we can detect DM subhalos with a significance of about $5\sigma$, that the size of extension would be roughly $0.8^{\circ}$, and that the significance of extension would be about $4\sigma$. 
   
   As for systematic uncertainties, we studied the case where our simulated data are randomized following Poisson statistics.
   Adding Poisson noise did not affect the results and the same conclusions as above hold true in case of counts randomization.
   Other systematics that can possibly alter the signal detection and extension reconstruction are, for example, a mismatch between the {\it true} Galactic diffuse model and the one used in the fit and/or
 the presence of unmodeled sources or background components close to the subhalo.
 The systematic uncertainty due to imperfection of Galactic diffuse modeling is alleviated when considering latitudes $|b|>20^{\circ}$ and energies $>1$ GeV.
 Considering the source count distribution of extragalactic sources recently derived in \cite{Marcotulli:2020fpm}, we can estimate about 1.0 source per deg$^2$ for fluxes above 100 MeV higher than $10^{-10}$ ph/cm$^2$/s. For fluxes of the order of $10^{-10}$ ph/cm$^2$/s the $TS$ for detection of a source is typically lower than 25. Assuming this number as an estimate of the density of extragalactic sources that shine below the {\it Fermi}-LAT detection threshold, we see that the presence of unmodeled and faint sources could be relevant for the search of DM subhalos, since there should be at least one faint extragalactic object in the innermost 1 deg$^2$ around the subhalo.
 
Although we do not address these systematics here, we expect them to be relevant in real data analyses and should be therefore properly taken care of when performing DM subhalo searches in real data.

Our analysis relies on subhalos having a cuspy density profile (NFW or Einasto) in agreement with the predictions of the cold DM scenario. If subhalos have cored profiles instead, they would be fainter and more susceptible to tidal effects, which would decrease their number and spatial extent. We expect this to decrease the detection significance associated to the extension, however the subhalo model we used is not designed to handle cored objects, and dedicated simulations are needed in order to correctly estimate the impact of subhalo distribution and statistics.

We stress that detecting \textsl{one} DM subhalo is a necessary condition for the discovery of DM. However, this may not be of course sufficient to attribute the signal to DM. To this end, it would be of interest, in future, to show what is the LAT sensitivity to the \textsl{simultaneous} detection of two or more subhalos. 
While we limit ourselves to the detectability of the brightest subhalo, we checked what is the statistics of $J$-factor and angular extension for the second- and third-brightest subhalos. 
We find that both of them have an extension comparable to that of the brightest subhalo and that the corresponding mean $J$-factor are less than $1\sigma$ away from $J_{\rm tot}^\star$. In particular, by rescaling our results for the mean $J$-factors ratios, we can estimate that, 
in order to detect the second- and the third-brightest subhalos with the same detection significance as the brightest one, we would need an increase of the cross section of a factor of 1.82 (1.70) and 2.63 (2.24) for the SL17-resilient (SL17-fragile) subhalo model.

We expect the general conclusions reached in the present work to apply also to searches for DM subhalo with the upcoming Cherenkov Telescope Array (CTA). 
CTA will be mostly sensitive to DM masses above 100 GeV.
In this DM mass range, given the significant improvement in angular resolution with respect to the LAT, CTA will provide a much better sensitivity to point-like and extended sources, and therefore improved perspectives for detection of the extension of DM subhalos. A quantitative estimate of such prospects is left for future analysis.

Finally, while our work focused on sensitivity predictions, we foresee application to real \Fermi-LAT data to look for {\it extended} DM subhalos, extend previous searches, and possibly set constraints on the DM parameter space~\cite{Inprep2021}.

\medskip

\begin{acknowledgments}
We warmly thank F. Donato and P. D. Serpico for careful reading of the manuscript and helpful discussions.
FC and MS acknowledge support by the Programme National Hautes \'Energies (PNHE) through the AO INSU 2019, grant ``DMSubG" (PI: F. Calore). 
Visits of MDM to LAPTh were supported by Université Savoie Mont-Blanc, grant ``DISE" (PI: F. Calore).
\end{acknowledgments}

\bibliography{extSH}

\end{document}